\documentclass[twocolumn]{svjour3}
\usepackage[english]{babel}
\usepackage{amsmath}
\usepackage{amssymb}
\usepackage{amstext}
\usepackage{graphicx}
\usepackage{mathptmx}

\newcommand{\ket}[1]{\left\vert #1\right\rangle}
\newcommand{\bra}[1]{\left\langle#1\right\vert}

\newcommand{\eps}{\varepsilon}
\newcommand{\be}{\begin{equation}}
\newcommand{\ee}{\end{equation}}
\newcommand{\tr}[1]{\mathrm{Tr} \left [ #1 \right ]}

\newcommand{\bea}{\begin{eqnarray}}
\newcommand{\eea}{\end{eqnarray}}

\newcommand{\ZZ}{\mathbb{Z}}
\newcommand{\RR}{\mathbb{R}}

\newcommand{\cA}{\mathcal{A}}
\newcommand{\cB}{\mathcal{B}}
\newcommand{\cH}{\mathcal{H}}

\begin{document}

\title{Charge transport through interfaces: a tight-binding toy model and its implications}

\author{B. A. Stickler \and W. P\"otz}
\institute{B.A. Stickler \and W. P\"otz
      \at
      Institute of Physics, Karl-Franzens Universit\"{a}t Graz, Austria.\\\email{benjamin.stickler@uni-graz.at}
      }
 
\maketitle

\begin{abstract}
With the help of a tight-binding (TB) electronic-structure toy model we investigate the matching of parameters across hetero-interfaces .   We demonstrate that the virtual crystal approximation, commonly employed  for this purpose, may not respect underlying symmetries of the electronic structure.  As an alternative approach we propose a method which is motivated by the matching of wave functions in continuous-space quantum mechanics. We show that this method obeys the required symmetries and can be applied in simple band to band transitions. Extension to multiple interfaces and to more sophisticated TB models is discussed.
\end{abstract}

\section{Introduction}

The modeling of quantum transport in nano-structured semiconductor devices is one of the major challenges in applied solid-state physics ~\cite{lake97,strahberger2000,boykin91,stovneng94,dicarlo2003,dicarlo94,ertler11,ertler10,ertler12,datta}.  In most cases a reliable quantum mechanical treatment of charge transport based on ab-initio electronic structure calculations seems to be impossible with state-of-the-art methods.  For many material classes even "ab-initio" methods require empirical input in order to reliably capture the  electronic structure in equilibrium ($T=0$).   Furthermore,  they generally cannot capture the non-linear-response regime ~\cite{bengone04,mavropoulos04}. We remark that these shortcomings may not apply to the method suggested by Brandbyge {\it et al.}~\cite{brandbyge02}.

As a resort, an empirical framework for the modeling of electronic devices under a finite bias has commonly been used.  The calculation is subdivided into three steps: (i) the electronic properties of (the components of) the  device are identified by means of computational or experimental efforts, (ii) the findings of step (i) are implemented into a model Hamiltonian which in step  (iii) is used in the  transport calculations, usually, with the inclusion of (self-consistent) corrections to account for the specific non-equilibrium situation.    
Typically, step (i) yields the (bulk) electronic structure of individual components of the device, as well as the band offset, for example, from a super-cell calculation.   In step (ii) this information is combined into a model Hamiltonian for the entire device.  Here one faces the problem of matching two or more regions with limited information regarding the interface.  In the simplest case, one may know the electronic structure of two bulk semiconductors and their relative  band alignment and be faced with the task to design an effective Hamiltonian for the study of charge transport across the hetero-interface.  

In the case of single-band electronic transport in the para\-bolic regime, an effective-mass model (EMM) may be appropriate. There is a plethora of analytic, as well as numerical, methods available to solve the associated  boundary value problem. In particular, if a linear voltage drop is assumed, Gundlach's method may be employed to solve the problem analytically~\cite{gundlach66}. However, the eigenvalues of the problem are substantially influenced by the boundary and matching conditions imposed upon the Schr\"{o}dinger equation. We note that the EMM may be viewed as a special one-band limit of the envelope-function approach. The matching strategy for the latter also faces problems when the overlaps of the Bloch functions for the two bulk materials to be joined are not known~\cite{poetz87}.

For more complex band structure situations, empirical tight-binding (TB) models provide a popular modeling tool ~\cite{dicarlo2003,stovneng94,datta,lake97,ertler10,ertler11,strahberger2000}.  In this case the TB parameters  have to be regarded as pure fitting parameters, ~\cite{vogl83,jancu98,starrost95,boykin04}, to be distinguished from the case where they stem from ab-initio calculations ~\cite{skriver,andersen98,tan2013}.  Within this model, transport calculations usually are  performed within the framework of non-equilibrium Green's functions (NEGFs) or the transfer matrix method ~\cite{datta,dicarlo2003}. 

 Besides its numerous benefits, such as conceptual simplicity and computational effectiveness, the empirical TB approach also comes with caveats.  Recently it has been argued by Tan {\it et al.} ~\cite{tan2013} that the reliability of transport calculations based on empirical TB parameters is rather questionable and, therefore, a direct mapping procedure based on ab-initio methods is preferable. Here, we shall address an even more serious disadvantage of commonly employed techniques in cases where the only information available consists of the bulk electronic structure and the band-offsets between the materials involved ~\cite{wei98,li09}.  A common approach to the modeling of the interface for binary compounds is the virtual crystal approximation (VCA) ~\cite{harrison,diventra,stovneng94,lake97,strahberger2000,boykin91,dicarlo2003}.  It is based on a simple (linear) interpolation of  TB  hopping and/or on-site parameters at the interface for establishing a matching between materials. 
 
 Within this work and utilizing a toy model we demonstrate  that the commonly used VCA for constructing the interface TB elements may not respect all symmetries of the underlying TB model and we propose an alternative approach which preserves these symmetries. It is important to remark that the method proposed by Brandbyge {\it et al.} ~\cite{brandbyge02} also employs an ad hoc approximation which is of the form of the VCA.  This paper is structured as follows: In Sec. \ref{sec:model} we define a rather simple model Hamiltonian and briefly discuss the NEGF method, as well as the EMM, as required for our purpose. In Sec. \ref{sec:dilemma} we demonstrate that the VCA is likely to lead to inconsistencies by investigating an artificial and a genuine interface. Finally, in Sec. \ref{sec:sol} we propose an alternative formulation of the interface matching problem and demonstrate that it respects the symmetries supplied by the input information. Conclusions are drawn in Sec. \ref{sec:conc}.

\section{The model} \label{sec:model}

We define a bulk toy model in order to formulate the matching problem and to demonstrate the inconsistencies which may arise. It is an infinite one-dimensional two component tight-binding chain with nearest-neighbor hopping only. Each element of the chain contains one orbital and the grid-spacing is given by $a > 0$. The Hamiltonian may be written as
\bea \label{eq:1}
H  &= & \sum_{l \sigma} \eps_\sigma \ket{l, \sigma} \bra{l, \sigma} \notag \\
 && + t_{12} \sum_l \left ( \ket{l, 2} \bra{l+1, 1} + \ket{l, 1} \bra{l, 2} \right ),
\eea
where $l \in \ZZ$ labels the unit cells, $\sigma \in \{ 1,2 \}$ labels the atoms within the unit cells and $\ket{l, \sigma}$ are the basis-kets localized at lattice point $(l,\sigma)$. Furthermore, $\eps_\sigma \in \RR$ are the onsite energies of atom $\sigma$ and $t_{12}$ is the hopping element. Please note that in writing Eq. \eqref{eq:1} we assume that $t_{12} \in \RR$ for reasons of simplicity. Let us define an onsite matrix $\eps$ and a hopping matrix $t$ via
\be \label{eq:2}
\eps = \left ( \begin{array}{c c}
                      \eps_1 & t_{12} \\
                       t_{12} & \eps_2
                     \end{array}
\right )
\ee
and
\be \label{eq:3}
t = \left ( \begin{array}{c c}
                   0 & 0 \\
                   t_{12} & 0
                  \end{array}
\right ).
\ee
The eigen-energies of the Hamiltonian \eqref{eq:1} can be expressed as
\be \label{eq:4}
E_{1,2} (k) = B \pm \sqrt{B^2 - A(k)},
\ee
where we introduce the wavenumber $k \in \left [ - \frac{\pi}{a}, \frac{\pi}{a} \right ]$ and define
\be \label{eq:5}
A(k) = \det \left [ { h}(k) \right ],
\ee
together with
\be \label{eq:6}
B = \frac{1}{2} \tr{ { h} (k) }.
\ee
Moreover, we define the Hamiltonian matrix $h(k)$ as the representation of the Hamiltonian $H$ in reciprocal space,
\be \label{eq:7}
{ h}(k) = \eps + t \exp ( i k a) + t^\dagger \exp ( - ik a).
\ee
Please note that we refer to the operator \eqref{eq:1} as the Hamiltonian while we denote the matrix $h(k)$ as the Hamiltonian matrix. The band structure defined by Eq. \eqref{eq:4} is completely determined by the set of TB parameters ${ \xi} = (\eps_1, \eps_2, t_{12} ) \in \RR^3$.  In order to emphasize this dependence we denote $E_n(k) \equiv E_n(k, { \xi})$, where $n = 1,2$.

We shall now consider two different bulk materials each characterized by a Hamiltonian of the form \eqref{eq:1} with TB parameters $\xi_L$ and $\xi_R$, respectively. We bring these two materials into contact and investigate the resulting heterostructure assuming that the band-offset is known and no further interface effects are taken into account. In what follows we present two common approaches to calculate the transmission function $T(E)$ through the interface for some particular energy $E$. Please note that we restrict our discussion to the case of zero bias for reasons of simplicity. The general arguments also apply to heterostructures under bias.

As a first technique we shall discuss the NEGF approach with the VCA. We introduce two Hamiltonians of the form Eq. \eqref{eq:1}, however, each defined only on one half axis, i.e. $l\in \ZZ^-$ and $l \in \ZZ^+$. Furthermore, we denote $H_{L}$ and $H_R$ the Hamiltonians defined on the semi-infinite domains of the left material $L$ ($l = -1,-2,\ldots$) and the right material $R$ ($l = 1,2,\ldots$). Diagonal elements of the TB Hamiltonians are shifted to comply with the band alignment. These two Hamiltonians are then connected with the help of an interface Hamiltonian $H_I$ such that the total Hamiltonian $\cH$ of the system is of the form
\be \label{eq:10}
\cH = H_L + H_I + H_R.
\ee
It is obvious that the particular choice of $H_I$ significantly influences the transport physics of the system, yet, in most applications its exact structure is unknown. In the particular case of two diatomic materials $AB$ and $CB$ which share the atomic constituent $B$, as for instance in the case of a GaAs / AlAs heterostructure~\cite{stovneng94}, a common approach is the virtual crystal approximation, for instance~\cite{harrison,diventra}, in which $H_I$ contains one element from the left chain, associated with $A$, while the onsite element of the common atom $B$ is averaged at the interface, i.e.
\be \label{eq:vca}
\eps_{2}^{I} = x \eps_{2}^L + (1 - x) \eps_{2}^R
\ee
where $x \in [0,1]$. Here, $\eps_2^{L/R}$ denote the onsite enery of atom $B$ according to the TB parameter sets $\xi_{L/R}$. The coupling to the left semi-infinite Hamiltonian is then described according to $t_{12}^L$, while the coupling to the right semi-infinite Hamiltonian is described by the hopping between atoms $B$ and $C$, i.e. $t_{12}^R$.

The transmission function $T(E)$ across the interface associated with Hamiltonian \eqref{eq:10} may, for instance, be calculated within the NEGF formalism, dividing the TB chain into three segments: the left semi-infinite lead consisting of the unit cells $l = -2,-3,\ldots$, the interface region referred to as system $S$ given by the unit cells $l = -1,0,1$,  and the right semi-infinite lead consisting of the cells $l = 2,3,\ldots$.  $T(E)$ is given by ~\cite{datta}
\be \label{eq:15}
T(E) = \tr{ \Gamma_R G_S^R \Gamma_L G_S^A},
\ee
where $\tr{ \cdot}$ denotes the operator trace, $\Gamma_{R / L}$ are the coupling functions to the right $(R)$ and the left $(L)$ semi-infinite leads, respectively, and $G_S^{R/A}$ denote the retarded and advanced system's Green's function. The latter are given by~\cite{datta}
\be \label{eq:16}
G_S^R = \left [ (E + i \eta) I - H_S - \Sigma_L^R - \Sigma_R^R \right ]^{-1},
\ee
and $G_S^A = \left ( G_S^R \right )^\dagger$ and $H_S$ is the system Hamiltonian. In Eq. \eqref{eq:16} $\eta > 0$ is a small parameter, $I$ is the identity, and $\Sigma^R_{L/R}$ is the retarded self-energy of the leads. The retarded self-energy of the leads is calculated from the surface Green's functions of the left and the right lead $G_{L / R}^R$, respectively, via~\cite{datta}
\be \label{eq:17}
\Sigma_{L}^R = { t^\dagger} G_{L}^R { t}, \qquad \text{and} \qquad \Sigma_{R}^R = { t} G_{R}^R { t^\dagger}
\ee
where we take into account that $t_{12} \in \RR$. The surface Green's functions, in general, can be determined recursively with the help of layer doubling, as suggested by Sancho {\it et al.} ~\cite{sancho85}. The coupling functions appearing in Eq. \eqref{eq:15} are given by
\be \label{eq:18}
\Gamma_{L/R} = i \left ( \Sigma_{L/R}^R - \Sigma_{L/R}^A \right ).
\ee

An entirely different approach is to calculate the transmission with the help of an EMM. We calculate the effective mass of the band of interest, say $n$, at $k = 0$ via
\be \label{eq:19}
m(\xi) = \hbar^2 \left [ \frac{\mathrm{d}^2 E_n(k, \xi)}{\mathrm{d} k^2} \right ]^{-1}_{k = 0}.
\ee
In what follows, we employ the notation
\be \label{eq:20}
m = \begin{cases}
     m_L & x \leq 0, \\
     m_R & x > 0,
    \end{cases}
\ee
where we replace the discrete index $l \in \ZZ$ by the continuous variable $x \in \RR$. Again, we partition the domain into two different regimes: (I) with $m = m_L$ for $x \leq 0$ and (II) with $m = m_R$ for $x > 0$ and solve the corresponding eigenvalue problem analytically. In particular, in region (I) we have
\be  \label{eq:21}
\psi''(x) + \frac{2 m_L}{\hbar^2} E \psi(x) = 0,
\ee
with the solution
\be \label{eq:22}
\psi(x) = \exp(ik_L x) + R \exp(-i k_L x).
\ee
Here, $R$ is the reflection amplitude and the wavenumber $k_L$ is given by
\be \label{eq:23}
k_L = \sqrt{\frac{2m_L}{\hbar^2} E}.
\ee
In similar fashion we obtain for regime (II) the solution
\be \label{eq:24}
\psi(x) = \tilde T \exp(i k_L x),
\ee
when no right-incident waves are considered. Here $\tilde T$ is the transmission amplitude and the wavenumber $k_R$ reads
\be \label{eq:25}
k_R = \sqrt{\frac{2 m_R}{\hbar^2} (E + \Delta)},
\ee
with $\Delta \in \RR$ the bandoffset. If we consider a linear voltage drop between two leads, one may solve the corresponding Schr\"{o}dinger equation analytically with the help of Airy functions ~\cite{gundlach66}. The numerical constants $R$ and $\tilde T$ are determined by the continuity conditions of the wave function and the current density.  These conditions read
 \begin{subequations} \label{eq:34}
 \bea
 1 + R & = & \tilde T, \\
 \frac{k_L}{m_L} \left ( 1 - R \right )  & = & \frac{k_R}{m_R} \tilde T.
 \eea
\end{subequations}
The transmission function $T(E)$ is then calculated via
\be \label{eq:35}
T(E) = \frac{k_R m_L}{k_L m_R} \vert \tilde T \vert^2.
\ee

A major benefit of the EMM is clearly its conceptual simplicity, but, in most cases the parabolic approximation may not be justified and the method of choice is the determination of $T(E)$ within a NEGF approach as discussed above. However, by writing Eq. \eqref{eq:10} for the total Hamiltonian $\cH$ we employ information which we actually do not have. Even worse, it turns out that the electronic structure is invariant under certain transformations of the TB parameters  and that these symmetries are destroyed by the particular choice of $H_I$ in the VCA \eqref{eq:vca}. Let us exemplify this dilemma within the next section in more detail.

\section{The dilemma} \label{sec:dilemma}

Let us define the operators $\cA$ and $\cB$ acting on vectors $\xi \in \RR^3$ via
\begin{subequations}
\be \label{eq:symop1}
\cA: (\eps_1,\eps_2,t_{12}) \mapsto (\eps_1,\eps_2,- t_{12}),
\ee
and
\be \label{eq:symop2}
\cB: (\eps_1,\eps_2,t_{12}) \mapsto (\eps_2,\eps_1,t_{12}).
\ee
\end{subequations}
Then we note from Eq. \eqref{eq:4} - \eqref{eq:7} that the electronic structure $E_n(k,\xi)$ obeys the invariances
\begin{subequations}
\be \label{eq:8}
E_n\left (k, \cA \xi  \right ) = E_n \left ( k,  \xi \right ),
\ee
and
\be \label{eq:9}
E_n \left ( k, \cB \xi \right )= E_n \left ( k, \xi \right ),
\ee
\end{subequations}
for all $\xi \in \RR^3$. Thus, the electronic structure described by the Hamiltonian Eq. \eqref{eq:1} is invariant under a change of the sign of the hopping parameter connecting the two atomic species, see Eq. \eqref{eq:8}. Furthermore, the electronic structure is also invariant under a relabeling of the atoms within the unit cell, as expressed in Eq. \eqref{eq:9}. This has the consequence that TB parameters are {\em not} uniquely determined by the band structure while the reversed statement is entirely true. Moreover, given a set of TB parameters we can immediately construct three further sets of parameters which yield completely identical bands by simply employing the invariances \eqref{eq:8} and \eqref{eq:9}.

In this light, the VCA for TB models may become problematic because one needs to unambiguously identify the associated atomic species, which according to Eq. \eqref{eq:9}  is not possible in many cases. In particular, $\eps_2(\cB \xi_L) = \eps_1(\xi_L)$. On the other hand, the EMM respects the symmetries Eqs. \eqref{eq:8} and \eqref{eq:9}, since the effective mass is solely based on the electronic structure, see Eq. \eqref{eq:19}.

In summary, if the only information available is the electronic structure $E_n(k)$ then the set of TB parameters defining the underlying Hamiltonian \eqref{eq:1} is not unique.  This is not problematic as long as one deduces from the Hamiltonian \eqref{eq:1} observables which too are invariant under a change in the TB parameters according to $\cA$ and $\cB$.  On the other hand, if the observables are not invariant under these transformations  one introduces an inconsistency because one utilizes information which is not contained in the electronic structure, as for instance, the sign of $t_{12}$. 

In what follows we investigate  this problematic in terms of the toy model in order to quantify the error.  

\subsection{An artificial heterostructure} \label{sec:arthet}

\begin{figure}
 \centering
\includegraphics[width = 70mm]{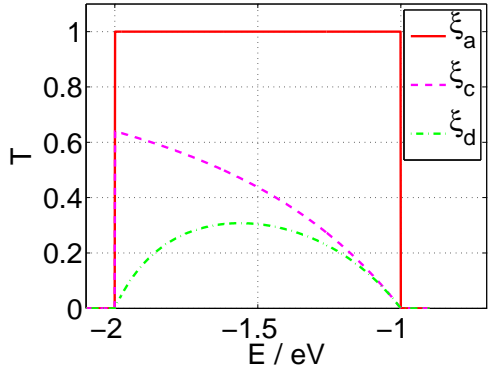}
 \caption{(Color online) The transmission function through the lower band resulting from an artificial interface as obtained by applying the symmetry operators $\cA$ and $\cB$ to $\xi$. Please note that we did not plot the transmission of $\xi_b$ because it is equivalent to the correct case $\xi_a$.} \label{fig:arthet1}
\end{figure}

For a first example we consider a homogeneous system with a bulk electronic structure described by a Hamiltonian of the form \eqref{eq:10}, i.e. we model a diatomic material comparable to, for instance, GaAs. 
The TB parameters $\xi = (2,-1,1)$ uniquely determine the electronic structure. Now an artificial interface between two semi-infinite linear chains is constructed by employing the symmetry operators $\cA$ and $\cB$ to the right chain. For the left semi-infinite chain we keep  $\xi_L = \xi$ , while for the right semi-infinite chain $\xi_R$ is one of the parameter sets: $\xi_a = \xi_L $, $\xi_b = \cA \xi_L$, $\xi_c = \cB \xi_L$  and $\xi_d = \cA \cB \xi_L$.  Since the transmission obtained with an EMM is invariant under these substitutions we shall  restrict the following discussion to the NEGF formalism with the VCA. The electronic structure, as well as the resulting transmission function for the lower band with $x = 0.5$, are illustrated in Fig. \ref{fig:arthet1}.

\begin{figure}
 \centering
\includegraphics[width = 70mm]{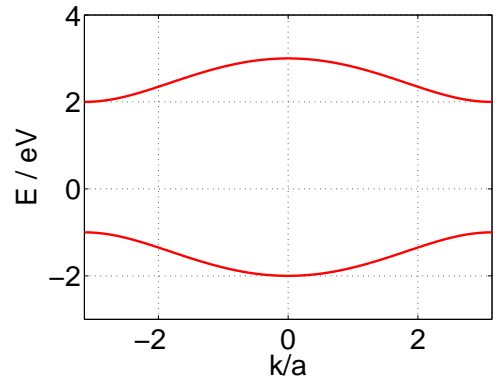}
 \caption{(Color online) Electronic structure of the toy model Eq. \eqref{eq:1} with TB parameters $\xi = (2, -1 ,1)$.} \label{fig:arbnds1}
\end{figure}

\begin{figure}
 \centering
\includegraphics[width = 70mm]{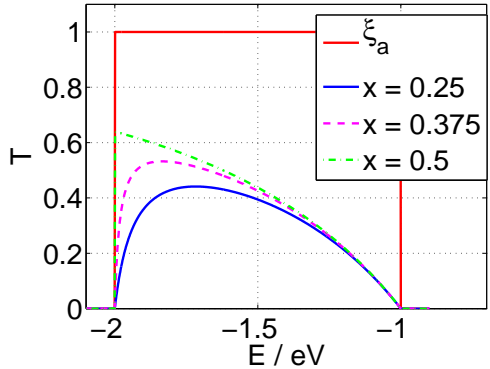}
 \caption{(Color online) The transmission function for the lower band resulting from an artificial interface with $\xi_c$ for different values of $x$.} \label{fig:arthet2}
\end{figure}

The transmission function resulting from the correct treatment of the material ($\xi_R = \xi_a$, red solid line) agrees with the result obtained with the help of $\xi_R = \xi_b$ and with that of an EMM. However, the transmission originating from an exchange of the parameters $1 \leftrightarrow 2$ according to $\cB$ strongly deviates from the original curve. The reason obviously stems from modeling an interface of the form $\cdots ABABBABA \cdots$ instead of $\cdots ABABABAB \cdots$. This example clearly illustrates a major problem in the TB treatment of interfaces with the help of the VCA. Moreover, we do not know which value of $x$ in the VCA \eqref{eq:vca} should give the correct result. We illustrated this dependency for the case $\xi_R = \xi_c$ in Fig. \ref{fig:arthet2} for three different values of $x$.

\subsection{A genuine heterostructure} \label{sec:truehet}

For the second example  we investigate  two truly different materials characterized by $\xi_L$ and $\xi_R$ and which have one atom in common, i.e. we are looking at an interface of the form $\cdots ABABCBCB \cdots$. According to the above discussion it is not possible to assign TB parameters to a particular atom within the unit cell as expressed in Eq. \eqref{eq:9}. This point is particularly crucial if the TB parameters are genuine fitting parameters stemming from the electronic structure solely, as in the case of empirical TB approaches~\cite{vogl83,jancu98,starrost95}. Then  it is not possible to unambiguously identify certain parameters with one atomic species in general.

\begin{figure}
 \centering
\includegraphics[width = 70mm]{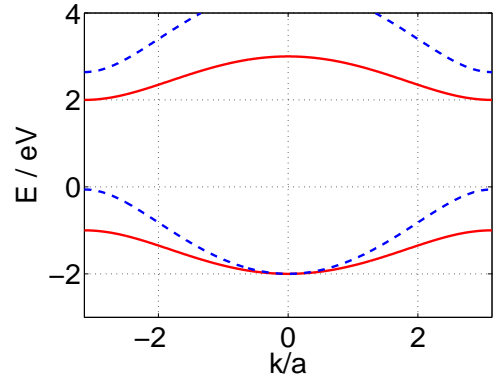}
 \caption{(Color online) Electronic structure according to the TB parameters $\xi_L = (2,-1,1)$ (red solid lines) and $\xi_R = (2.6398,-0.0602,1.5)$ (blue dashed lines).} \label{fig:trueband}
\end{figure}

\begin{figure}
\includegraphics[width = 70mm]{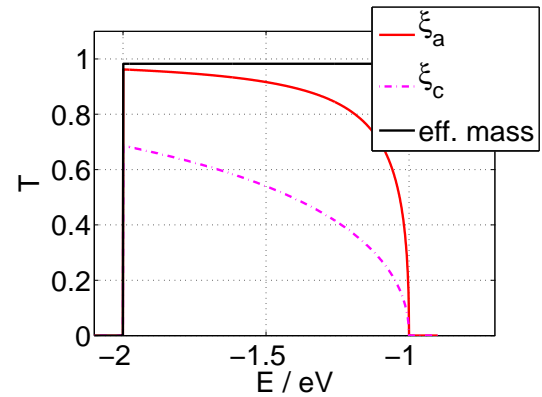}
 \caption{(Color online) Transmission $T(E)$ for the composite material with the NEGF technique with parameters $\xi_a$ (red solid), $\xi_c$ (magenta dash dotted) and for the EMM (black solid line). The curves for $\xi_b$ and $\xi_d$ have been omitted since they are equivalent to $\xi_a$ and $\xi_c$, respectively.} \label{fig:truetrans1}
\end{figure}

The two electronic structures depicted in Fig. \ref{fig:trueband}, respectively, correspond to the TB parameters  $\xi_L = (2,-1,1)$, for $AB$,  and  $\xi = (2.6398,-0.0602,1.5)$, for  material $BC$. Again, $\xi_R$ takes is equivalent to one of the TB parameters $\xi_a = \xi$, $\xi_b = \cA \xi$, $\xi_c = \cB \xi$ and $\xi_d = \cA \cB \xi$. The resulting transmission functions for the lower band are depicted in Fig. \ref{fig:truetrans1}.

In summary, for a simple model system we have identified the reasons  why the VCA does not yield satisfactory results. The problem is that (i) an arbitrary mixing parameter $x$ is introduced at the interface, and (ii)  the symmetry of the electronic structure under the operators $\cA$ and $\cB$ acting on $\xi$, Eqs. \eqref{eq:8} and \eqref{eq:9}, is not respected within the VCA.  It is clear that ambiguities grow dramatically with increasing complexity of the TB model.  
A determination of the transmission function within the EMM respects this symmetry, however, the EMM itself often is too crude for an electronic structure model.  In the EMM the preservation of symmetry in the transmission function is a result of the matching condition at the interface, see Sec. \ref{sec:model}.  In what follows we propose a comparable approach for TB models as a possible method for avoiding ambiguities in the interface problem.

\section{The interface matching problem - a possible solution} \label{sec:sol}

Within this section we propose a discrete matching approach for TB models which is adopted from the matching procedure within the EMM. The main procedure consists in solving the bulk problem for the two different materials separately and matching the wave functions, as well as the current between unit cells, at the interface. Such an approach has two main advantages: (i) it does not introduce arbitrary parameters at the interface and (ii) it respects the symmetries of the electronic structure, e.g.  under variation of $\xi$ according to $\cA$ and $\cB$, Eqs. \eqref{eq:8} and \eqref{eq:9}, for the case of the TB Hamiltonian  \eqref{eq:1}.  However, this approach still cannot account for interfacial effects as long as no additional information is made available. It has, therefore, to be regarded as a best solution based on the information given. We shall first illustrate the procedure for the matching of two single-atomic linear TB chains in subsection \ref{sec:match1} since it allows an 
entirely 
analytic exposition of the method. Subsequently, the method is generalized and then applied to the toy model in subsection \ref{sec:truehet2}.

The assumption of an abrupt interface between to semi-infinite crystals inevitably represents an approximation which leads to a loss of information regarding microscopic details at the interface.  
In an empirical TB model parameters should become position dependent near the interface.  However, the mapping of ab-initio band structure calculations onto empirical TB models numerically is rather intractable (due to  supercell size) and one may resort to an approach based on stationary scattering theory.  Considering a quasi-1D system, such as a heterostructure, a stationary solution 
for given energy $E$ with in-asymptote $\ket{n,k_{||},k_i} $, with band index $n$  and parallel k-vector $k_{||}$ ,   may be written as 

\be
 \ket{E} = \begin{cases}
            \ket{ n,k_{||},k_i}  + \sum_{m,j}^{\mbox{(out)}} r_{n,i;m,j}(E,k_{||})  \ket{m,k_{||},k_j}  & \text{ for } z<0, \\
            \sum_{m',j'}^{\mbox{(out)}} t_{n,i;m',j'}(E,k_{||}) \ket{m',k_{||},k_{j'}} & \text{ for } z>0 \label{zg0} ,
           \end{cases}
\ee
when the interface is positioned at "$z=0$".  The sum is over the out-channels for which $E=E_m(k_{||},k_j)=E_{m'}(k_{||},k_{j'})$.  In- and out-channels, respectively, are identified as having group velocity $z$-components  towards the interface and away from the interface.  In general,  degeneracy may imply more than two out-channels $m,k_j$ for each in-channel.  Unitarity of the S matrix and possibly other symmetries, such as time-reversal invariance, reduce the number of independent elements $r_{n,i;m,j}(E,k_{||})$ and $t_{n,i;m',j'}(E,k_{||})$ but are not sufficient to uniquely specify them if the interface potential at $z\approx 0$ is unknown. 

Let $N_L$ and $N_R$ denote the number of available in-channels on the left- and the right-hand side of the interface at a given energy $E$. Under time-reversal symmetry, we also have $N_L$ and $N_R$ out-channels on each side. Then, unitarity of the $S$ matrix leads to $(N_L + N_R)^2$ conditions for $2 (N_L + N_R)^2$ unknowns. For $N_L = N_R = 1$, i.e., one out-channel on each side, one is left with two unknowns.  For this case, and this represents our suggested alternative for an ad-hoc VCA,  the condition of continuity in charge and current density across the interface determine the transmission and reflection at the hetero-interface, in analogy to the effective-mass case and a finite potential step.
For higher degeneracy, additional information is necessary to determine all of the $S$ matrix elements.  

\subsection{Tight-binding interface matching for single-atomic chain} \label{sec:match1}

Two semi-infinite single-atomic linear  TB chains are to be connected at the grid point $l = 0$. The connection between these two chains is not established by introducing an ad-hoc  hopping parameter $\tilde t$ across the interface, as for instance suggested by Harrison~\cite{harrison}, but rather by {\em matching} the left and right solutions at the interface. For this purpose, the left- and right-semi-infinite TB Hamiltonian both are extended to site $l = 0$.  Specifically we propose the matching conditions
\begin{subequations} \label{eq:cond}
\be \label{eq:cond1}
\vert \psi_L(0) \vert^2 = \vert \psi_R(0) \vert^2,
\ee
and
\be \label{eq:cond2}
J_{-1/2}^L = J_{+1/2}^R.
\ee
\end{subequations}
Here, $\psi_{L/R}$ are the wave functions at site $l = 0$ expressed respectively,  in terms of the  left ($L$) and right ($R$)  TB  orbitals. The second condition  \eqref{eq:cond2} guarantees stationarity of the solution, i.e. the current flowing between the grid points $l = -1$ and $l = 0$ in the left chain equals the current flowing between $l = 0$ and $l = 1$ in the right chain.  While the matching conditions \eqref{eq:34} can be derived from the Schr\"{o}dinger equation, these  two conditions \eqref{eq:cond} may be viewed as a (stationary version of the) particle continuity equation in terms of TB orbitals. Condition \eqref{eq:cond1} ensures that the probability for finding a particle at the interface is the same when approaching the interface from left or right:  at the interface this probability can be expressed either in terms of the $L$  or $R$  basis functions.  In the parabolic regime above conditions \eqref{eq:cond}  become exact (i.e., agree with the ones for the Schr\"{o}dinger equation) and 
coincide with Eqs. \eqref{eq:34}.

The isolated Hamiltonian of the $L$ and $R$ single-particle TB chain, respectively, is denoted by $H_{L}$ and $H_R$ with onsite energies $\eps_{L}$, $\eps_R$ and hopping elements $t_{L}$, $t_R$, respectively. The corresponding energy bands are
\be \label{eq:dispersion}
E_{L/R}(k) = \eps_{L/R} + 2 t_{L/R} \cos(a_{L/R} k).
\ee
For both bulk Hamiltonians, we write the eigenfunctions as linear combination of localized states $\ket{l_{L/R}}$
\be
\ket{\Psi_{L/R}} = \sum_l a_l^{L/R} \ket{l_{L/R}},
\ee
where the general form of the coefficients $a_l^{L/R}$ for given $k$ is well known to be of the form
\be
a_l^{L/R}(k) = A_{L/R} \exp ( i k l a ) + B_{L/R} \exp ( - i k l a).
\ee
We assume for simplicity that the lattice constants coincide $a_L = a_R \equiv a$. In analogy to Eq. \eqref{eq:22} we choose
\be \label{eq:coff1}
A_L = 1, \quad B_L = R, \quad A_R = T, \quad \text{and} \quad B_R = 0.
\ee
For  given energy $E$, condition \eqref{eq:cond1} reads 
\be
\vert a_0^L[k_L(E)] \vert^2 = \vert a_0^R[k_R(E)]\vert^2,
\ee
where we define the wave numbers $k_{L/R}(E) = \vert E_{L / R}^{-1}(E) \vert$ with the help of the inverse of Eq. \eqref{eq:dispersion}. With  Eq. \eqref{eq:coff1} we rewrite this condition as
\be\label{eq:condeq1}
\vert 1 + R(E) \vert^2 = \vert T(E) \vert ^2.
\ee
We calculate the current from lattice point $-1$ to lattice point $0$ via
\bea
J_{-1/2} & = & \frac{i}{\hbar} \left [ (a^L_0)^* t_L a_{-1}^L - i (a_{-1}^L)^* t_L a_0^L \right ]\notag \\
 &= & - \frac{2}{\hbar} t_L ( \vert R \vert^2 - 1 ) \sin(k_L a),
\eea
and from $0$ to $1$ as
\be
J_{+1/2} =\frac{2}{\hbar} \vert T \vert^2 t_R \sin(k_Ra).
\ee
Furthermore, it is convenient to define the velocities
\be
v_{L/R}(k) = \frac{1}{\hbar}\frac{\mathrm{d}}{\mathrm{d} k} E_{L/R}(k) = - \frac{2}{\hbar} a t_{L/R} \sin (k a),
\ee
and rewrite condition \eqref{eq:cond2} as
\be \label{eq:condeq2}
\vert R(E) \vert^2 + \frac{v_R[k_R(E)]}{v_L[k_L(E)]} \vert T(E) \vert^2 = 1.
\ee
Hence, we solve the coupled equations \eqref{eq:condeq1} and \eqref{eq:condeq2}. Under the assumption that $R,T \in \RR$ these equations are easily solved to give
\begin{subequations}
\be
T(E) = \frac{2 v_L(k_L)}{v_L(k_L) + v_R(k_R)},
\ee
and
\be
R(E) = \frac{v_L(k_L) - v_R(k_R)}{v_L(k_L) + v_R(k_R)}.
\ee
\end{subequations}
Clearly, the quantities of physical interest are
\begin{subequations} \label{eq:coff2}
\be
\frac{v_L(k_L)}{v_R(k_R)} \vert T(E) \vert^2 = \frac{4 v_R(k_R)v_L(k_L)}{[v_L(k_L) + v_R(k_R)]^2},
\ee
and
\be
\vert R(E) \vert^2 = \left ( \frac{v_L(k_L) - v_R(k_R)}{v_L(k_L)+v_R(k_R)} \right )^2.
\ee
\end{subequations}

\begin{figure}
 \centering
\includegraphics[width = 70mm]{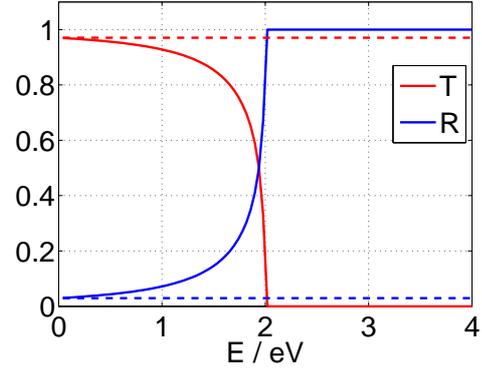}
\caption{(Color online) Reflection and transmission coefficient through the interface between two semi-infinite single atom TB chains as obtained with the matching method (solid lines) and with the EMM (dashed lines).}\label{fig:model1}
\end{figure}

\begin{figure}
 \centering
\includegraphics[width =70 mm]{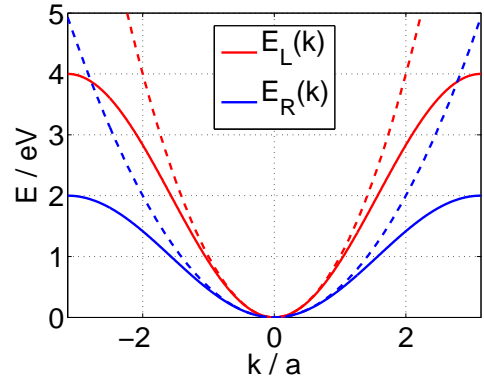}
 \caption{(Color online) Electronic structures of the two isolated semi-infinite chains (solid lines) and parabolic approximations (dashed lines).}\label{fig:model2}
\end{figure}

The transmission and reflection as a function of energy $E$ can be obtained by inverting Eq. \eqref{eq:dispersion}, however, it has to be kept in mind that, due to our choice of Eq. \eqref{eq:coff1}, one has to take the branch $v_L(k_L), v_R(k_R) \geq 0$. In Fig. \ref{fig:model1} we illustrate the transmission and reflection coefficients Eq. \eqref{eq:coff2} for $\eps_{L} = 2$, $\eps_R = 1$, $t_L = -1$, $t_R = -0.5$ in comparison with the result obtained with an EMM. The corresponding electronic structures Eq. \eqref{eq:dispersion} together with the parabolic approximations are illustrated in Fig. \ref{fig:model2}.

From Fig. \ref{fig:model1} we observe that the reflection and the transmission coefficients from the left to the right TB chain are well approximated by the EMM only for energies near to the band minimum. This is in accordance with the effective mass approximation, see Fig. \ref{fig:model2}.  Moreover, we note that the reflection correctly approaches $1$ as the energy $E$ approaches $2$ eV, see Fig. \ref{fig:model1}. This is due to the reduced band width of the electronic structure of the right semi-infinite chain, see Fig. \ref{fig:model2}.  The transmission is maximal for low energies.

In what follows we shall formulate the matching method for more general TB Hamiltonians and then apply it to the diatomic TB chain model discussed in subsection \ref{sec:truehet}.

\subsection{The genuine heterostructure revisited} \label{sec:truehet2}


Let us begin with a brief review of some important properties of general nearest neighbor TB models. The  Hamiltonian $H$ is of the form
\bea \label{eq:hamgen}
H & = & \sum_{l\sigma} \eps_\sigma \ket{l\sigma}\bra{l \sigma} \notag \\
 && + \sum_{l l' \sigma \sigma'} t_{l l'}^{\sigma \sigma'} \ket{l \sigma} \bra{l' \sigma'}
\eea
where $l$ and $l'$ label the unit cell and $\sigma=1,\ldots, N$ is an additional index. This additional index might include, for instance, the atoms in the unit cells as well as their orbitals and the parallel momentum $k_\|$, if the Hamiltonian \eqref{eq:hamgen} is derived from a three-dimensional TB model by partial Wannier transformation ~\cite{stovneng94}. The hopping elements $t_{ll'}^{\sigma\sigma'}$ couple only up to neighboring unit cells.    This is a valid assumption for any finite-range TB model since the unit cell can be chosen as large as necessary ~\cite{schulman83}. We expand the wavefunction as
\be
\ket{\psi} = \sum_{l \sigma} c_{l \sigma} \ket{l \sigma} \equiv \sum_l c_l \cdot \ket{l},
\ee
where in the very last step we introduce the $N$-component vectors $c_l = \{ c_{l\sigma} \}$ and a "vector ket" $\ket{l} = \{ \ket{l\sigma} \}$. Bloch's theorem allows one to extract the space dependence of the vector $c_l$ at a given energy $E$ in form of a phase factor
\bea
c_l(E) & = & c[k(E)] \exp \left [ i k(E) l a \right ] \notag \\
 && + c[-k(E)] \exp \left [ - i k(E) l a \right ],
\eea
where $k (E)$ is the inverse of the electronic structure $E=E_n(k)$ for a given energy $E$ and $c(k)$ is the eigenvector of the Hamiltonian matrix $h(k)$ with eigenvalue $E$. The Hamiltonian matrix $h(k)$ reads
\be
h(k) = \eps + \exp \left ( i k a \right ) t + \exp \left ( - i k a \right ) t^\dagger,
\ee
where we define an onsite matrix $\eps$ with matrix elements $\eps_\sigma \delta_{\sigma \sigma'} + t_{ll}^{\sigma \sigma'}$ and a coupling matrix $t$ with elements $t_{l l+1}^{\sigma \sigma'}$, respectively. Please note that the Hamiltonian matrix Eq. \eqref{eq:7} is a special case of this general form. Furthermore, the current $J_{l + 1/2}$ between unit cells $l$ and $l + 1$ can be written as
\be
J_{l+ 1 / 2} = \frac{i}{\hbar} \left ( c_l^\dagger t c_{l+1} - c_{l+1}^\dagger t^\dagger c_l \right ).
\ee

Hence, the general matching conditions for given energy $E$ read
\begin{subequations} \label{eq:match2}
\be
\left \vert c_L[k_L(E)] + c_L[-k_L(E)] R \right \vert^2 = \left \vert T c_R [ k_R(E)] \right \vert^2,
\ee
and
\bea
J^L_{-1 / 2} = J_{+1/2}^R,
\eea
where we use
\be
J_{-1/2}^L = \frac{i}{\hbar}  \left [ \left ( c^L_{-1} \right )^\dagger t_L c_{0}^L - \left ( c_{0}^L \right )^\dagger t_L^\dagger c_{-1}^L \right ],
\ee
with
\bea
c_l^L & = & c_L [ k_L(E) ] \exp \left [ i k_L(E) l a_L \right ] \notag \\
 && + R c_L [ k_L(E) ] \exp \left [- i k_L(E) l a_L \right ],
\eea
and
\be
J_{+1/2}^R = \frac{i}{\hbar}  \left [ \left ( c^R_{0} \right )^\dagger t_R c_{1}^R - \left ( c_{1}^R \right )^\dagger t_R^\dagger c_0^R \right ],
\ee
with
\be
c_l^R = T c_R [ k_R(E) ] \exp \left [ i k_R(E) l a_R  \right ].
\ee
\end{subequations}
Here, $c_{L/R}[k_{L/R}(E)]$ are the eigenvectors to the Hamiltonian matrices $h_{L/R}(k)$ with eigenenergies $E$, which are of the form Eq. \eqref{eq:7}, however, corresponding to the Hamiltonians $H_L$ and $H_R$, respectively. Moreover, with $k_{L/R}(E)$ for a given energy $E$ we define the inverse of the dispersion relations $E_n^L(k)$ and $E_n^R(k)$, respectively. Finally, $a_{L/R}$ denote the lattice constants of the two materials. Note that these matching conditions respect time-reversal symmetry.   

\begin{figure}
 \centering
\includegraphics[width = 70mm]{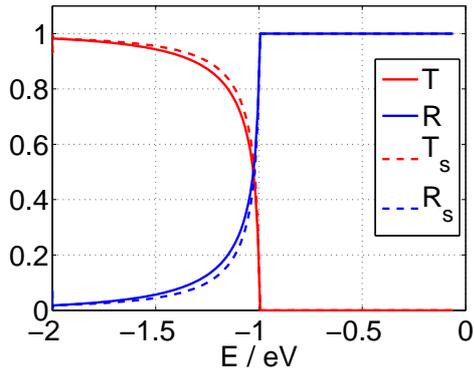}
 \caption{(Color online) Transmission and reflection coefficient through the interface between two semi-infinite diatomic TB chains as obtained with the matching method (solid lines) and as obtained when employing relations \eqref{eq:coff2} (dashed lines).} \label{fig:transdemo}
\end{figure}

\begin{figure}
 \centering
\includegraphics[width = 70mm]{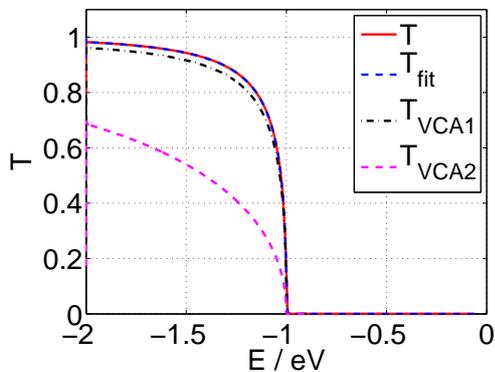}
 \caption{(Color online) Transmission coefficient throught the interface between two semi-infinite diatomic TB chains as obtained with the NEGF formalism where the hopping has been fitted to the result of the matching method, (solid red and dashed blue lines). Furthermore, we present the solutions obtained with the VCA in Sec. \ref{sec:truehet}. } \label{fig:truehetfit}
\end{figure}

We solve the above equations \eqref{eq:match2} for the problem discussed in Sec. \ref{sec:truehet}. The resulting transmission and reflection functions are plotted in Fig. \ref{fig:transdemo}, in comparison to the result obtained with the help of Eqs. \eqref{eq:coff2}. It is very important to notice that the result depicted in Fig. \ref{fig:transdemo} is {\em independent} of the actual choice of the parameter sets $\xi_L$ and $\xi_R$, i.e. the transmission function is independent under the action of the symmetry operators $\cA$ and $\cB$. This follows from the matching conditions \eqref{eq:coff2}, since all quantities entering these equations are solely determined by the bulk Hamiltonians $H_{L/R}$.  In Fig. \ref{fig:truehetfit} we illustrate the transmission through the diatomic TB heterostructure discussed in Sec. \ref{sec:arthet} in comparison with the results obtained with the help of the virtual crystal approximation, see Sec. \ref{sec:arthet}.

The  transmission and reflection function versus energy computed in this fashion may subsequently be mapped back onto a TB model. This can be achieved by constructing a Hamiltonian $\tilde \cH$ of the form
\be
\tilde \cH = H_L + V(E) + H_R,
\ee
where one introduces an energy dependent coupling Hamiltonian $V(E)$ between the two chains.  Following the procedure outlined in Sec. \ref{sec:model} to obtain the transmission $T(E)$ with the help of a Green's function approach,  the elements of the hopping matrix $V(E)$ is then determined in such a way that the transmission obtained by solving Eqs. \eqref{eq:coff2} is reproduced. Hence, $V(E)$ will depend on the particular choice of $\xi_{L}$ and $\xi_R$, however, is determined in such a way that it reproduces a parameter-independent result, such as the transmission $T(E)$.   Moreover, as long as the voltage drop at  the interface is small, $V(E)$  may provide a good approximation for the heterostructure under bias.  An extension to multiple interfaces of this type is straight-forward.

This method works as long as to every in-channel Bloch state there is only one out-channel state, respectively, to the left and to the right.  In other words, as long as one deals with the problem of matching a doubly degenerate band in one material with another one in the other.  The proposed matching conditions \eqref{eq:match2} alone are not sufficient to tackle the case of multi-channel scattering where, for given in-channel $E,n,k_{||}$ (energy, band index, and k-parallel) the degenerate out-channel states lie in different energy bands $n$, or in the same band $n$ when a degeneracy of greater than two is present. A unique solution cannot be obtained without further assumptions. This is most easily observed by inspection of a situation with one in-channel on the left hand side $i$ and two out-channels on the right hand side. In this case the wave function to the right of the interface is given by a linear combination of the first and the second channel states with weights $t_{i,1}(E)$ and $t_{i,2}(E)$  
in Eq.  \eqref{zg0}. It is clear, that the matching conditions Eqs. \eqref{eq:match2} may be solved only if the ratio $t_{i,1}(E)/t_{i,2}(E)$ is known.  An estimate for this ratio might be obtained by "blanking off" one out-channel at a time and determining the two transmission functions at a time.  However,  in this procedure, phase information is lost.  It should also be mentioned that in many multichannel problems, there may be  dominant out-channels (those which couple dominantly with a given in-channel). This may  be used to reduce the problem in an approximation  to a tractable $2$ by $2$ form of one out-channel on each side of the interface.  In addition, of course, empirical input from experiment may be helpful in the fitting procedure.

\section{Conclusions} \label{sec:conc}

We have studied the electron  transmission through a  hetero-interface as modeled by the linking of  two diatomic single-orbital TB chains. The two separate TB chains are characterized solely by their bulk properties while no particular information about the interface, other than the relative band alignment,  is available. For this example we demonstrate that the commonly employed VCA does not respect the underlying symmetries of the electronic structure in relation to the TB model used in the fitting procedure.   Commonly, however, the latter is the key input information available.  Hence, the VCA  introduces an arbitrary error which is hard to estimate.  As a remedy to this ambiguity we suggest a matching method of the wave function and current density which is motivated by the continuity relations of continuous space quantum mechanics. In particular, the obtained transmission functions respect the symmetries of the band structures associated with the TB model  and can therefore be regarded as a best 
result under available information.   The transmission function determined  with the help of the proposed matching method can subsequently be used to  construction of a hopping Hamiltonian which reproduces this transmission function. Thus, multiple interfaces can be treated in this fashion.

We stress that the  proposed matching method does not give a physically complete description of the interface. In fact, this is not possible since the required information regarding the coupling matrix at the interface which enters the total Hamiltonian $\cH$ is considered to be unknown. Hence, the approach discussed here is no substitute for a full (microscopic) study of the hetero-interface.  But, since such a study is almost always based on large supercell calculations, one is confronted with the problem of fitting numerous bands which, in most cases,  practically is not feasible.

The matching method is applicable to the case of one out-channel on each side (per in-channel). If more in- or out-channels are present, a simple matching technique cannot yield the channel resolved transmission  since the problem is underdetermined (since the Hamiltonian at the interface is unknown).  However, if the transmission ratio between all available out-channels is known, the proposed method can still be employed in order to obtain the total transmission function.

\section{Acknowledgments}

This work was supported financially by FWF project P221290-N16.

\bibliographystyle{unsrt}
\bibliography{report}

\end{document}